\documentclass{article}
\usepackage[utf8]{inputenc}
\usepackage{subcaption}
\usepackage{multirow}
\usepackage{graphicx}
\usepackage{marvosym}
\usepackage{paralist}
\usepackage{hyperref}
\usepackage{amsmath}
\usepackage{authblk}
\usepackage{colortbl}
\usepackage{xcolor}
\usepackage[style=numeric]{biblatex}
\usepackage{graphicx}
\usepackage{url}
\usepackage{listings}
\newcommand{\furl}[1]{\footnote{\scriptsize \url{#1}}} 
\usepackage[style=numeric]{biblatex}
\bibliography{references}
\definecolor{Gray}{gray}{0.85}
\definecolor{LightCyan}{rgb}{0.88,1,1}

\newcolumntype{a}{>{\columncolor{Gray}}c}
\newcolumntype{b}{>{\columncolor{white}}c}

\begin{document}
\title{Managing Knowledge in  Energy Data Spaces}
\author[1,2]{Valentina Janev}
\author[3,4]{Maria-Esther Vidal}
\author[3,4]{Kemele Endris}
\author[1,2]{Dea Pujić}
\affil[1]{The Mihajlo Pupin Institute, Serbia}
\affil[2]{University of Belgrade, Serbia}
\affil[3]{L3S Research Center, Leibniz University of Hannover, Germany}
\affil[4]{TIB Leibniz Information Centre for Science and Technology, Hannover, Germany}
\affil[ ]{\normalsize\texttt{valentina.janev@institutepupin.com, maria.vidal@tib.eu}}
\date{}
\maketitle

\paragraph{\bf Abstract}
Data in the energy domain grows at unprecedented rates and is usually generated by heterogeneous energy systems.
Despite the great potential that big data-driven technologies can bring to the energy sector, general adoption is still lagging. Several challenges related to controlled data exchange and data integration are still not wholly achieved. As a result, fragmented applications are developed against energy data silos, and data exchange is limited to few applications. 
In this paper, we analyze the challenges and requirements related to energy-related data applications. We also evaluate the use of Energy Data Ecosystems (EDEs) as data-driven infrastructures to overcome the current limitations of fragmented energy applications. EDEs are inspired by the International Data Space (IDS) initiative launched in Germany at the end of 2014 with an overall objective to take both the development and use of the IDS reference architecture model to a European/global level. The reference architecture model consists of four architectures related to business, security, data and service, and software aspects. This paper illustrates the applicability of EDEs and IDS reference architecture in real-world scenarios from the energy sector.
The analyzed scenario is positioned in the context of the EU-funded H2020 project PLATOON. \paragraph{\bf Keywords}
Data Integration Systems, Energy Big Data, Knowledge Graphs, Data Exchange, Semantic Interoperability
%\begin{CCSXML}
%<ccs2012>
%   <concept_id>10002951.10003260.10003309.10003315</concept_id>
%   <concept_desc>Information systems~Semantic web description languages</concept_desc>
%    <concept_significance>500</concept_significance>
%    <concept_id>10002951.10002952.10002953.10010146</concept_id>
%    <concept_desc>Information systems~Graph-based database models</concept_desc>
%</ccs2012>
%\end{CCSXML}

%\ccsdesc[500]{Information systems~Semantic web description languages}
%\ccsdesc[500]{Information systems~Graph-based database models}

\section{Introduction}
Big data is recognized as a relevant asset, and big data-driven applications are increasingly devised in several domains and disciplines. 
%Big data has been widely and successfully adopted in several fields such as health care, retail, and manufacturing. 
In the energy sector, big data has been presented as digital technology to understand how energy is produced and consumed and how these patterns may impact in our lives and economies~\cite{Forbes}. Despite recognized as crucial applications for efficiently generating and consuming energy, Big data applications in the energy domain are still underdeveloped and fragmented. Energy big data is collected from heterogeneous data sources which include wind farms, solar systems, conventional power plants, cooling, heating, and lighting systems as well as smart grids. They represent measurements in different domains, e.g., energy consumption, energy generation, system outages, failures, weather, and energy transmission. Moreover, these data sources are characterized by the dominant Big Data dimensions, i.e., volume, velocity, variety, veracity, and value. Furthermore, interoperability and heterogeneity are usually caused by the various representations and interpretations of the data ingested from the data sources. These results put in perspective data complexity issues that need to be tackled in the energy sector. This paper states the requirements to be fulfilled during data sharing and integration to scale up large data sets and solve heterogeneity and quality issues. 

 Data interoperability is defined as the process of providing uniform access to a set of distributed (or decentralized), autonomous, and heterogeneous data sources~\cite{doan2012principles}. 
 Data integration systems (DIS) integrates data sets; they provide a global schema (also known as mediated schema) to provide a unified view of all data available in different integrated data sources. DISs can produce a materialized version of a data warehouse of the integrated data sources. The unified schema serves not only to explicitly define the underlying data elements (thus achieving syntactic interoperability), but also to assign unambiguous, shared meaning to processed data, i.e., to reach semantic interoperability. Semantic interoperability has been a key consideration in information systems design in the last two decades and its importance has been widely recognised~\cite{DAVIES2020102426}. 
 %However, predominant big data dimensions (i.e., volume, variety, velocity, and veracity) negatively impact on the development of scalable and efficient DIS, in particular, in the energy domain. 

Several initiatives have been proposed to develop effective and scalable semantic interoperability towards data spaces. 
In the context of the European Data Strategy~\cite{EDS} and the proposed Regulation on European Data Governance~\cite{DataAct}, a vision has been created for trusted data intermediaries for B2B data sharing and common European data spaces in crucial sectors such as health, environment, energy, agriculture, mobility, finance, manufacturing, public administration and skills. The International Data Space (IDS) is another initiative to enable controlled data exchange and integration~\cite{BaderPMTQMABILG20}. IDS proposes various standards, technologies, and governance models to facilitate secure and standardized data exchange and integration. Moreover, IDS provides building blocks for the development of data-driven services, rs is guaranteed. 
Lastly, Data Ecosystems~\cite{capiello2020data,oliveira2018data} are infrastructures that allow for data exchange across different stakeholders; they are equipped with data integration techniques and data management methods to preserve data privacy and security. DEs facilitate the creation of data markets for ensuring competitiveness and data sovereignty. 

This paper analyzes the requirements for energy data exchange and illustrate with a real-world use case, interoperability issues that may exist across energy data sources. Furthermore,  DEs for energy data managements are presented as referential architectures to addressed challenges of the energy sector. 

The paper begins with preliminaries related to recent data sharing initiatives in Europe (\autoref{sec:prelim}). This is followed by  a motivation scenario from the energy sector in \autoref{sec:example}. Our approach for Big Data Management and Analytics in energy domain is introduced in \autoref{sec:approach}. Section \ref{sec:proof} presents proof-of-concept results and the observed outcomes are discussed in \autoref{sec:discussion}.
Related work are analyzed in \autoref{sec:related}, and \autoref{sec:conclusion} summarizes our conclusions and presents an outlook to our future work. 

 \begin{figure*}[t!]
     \centering
     \includegraphics[width=1.0\columnwidth]{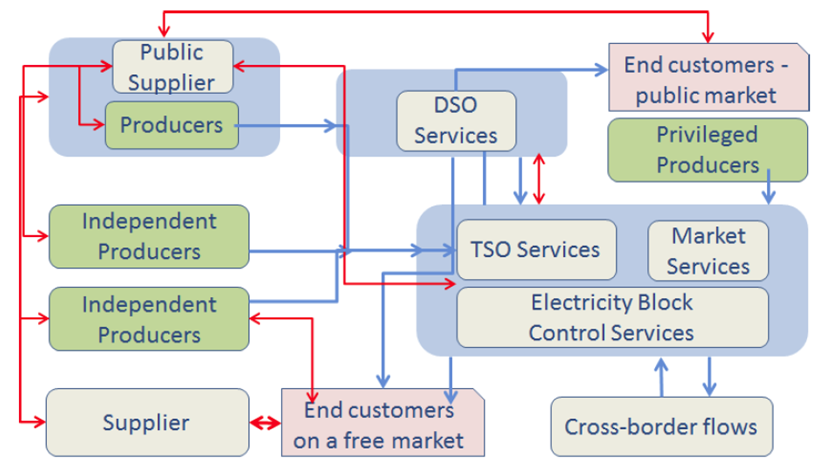} 
     \caption{Electricity Market - Information flows}
     \label{fig:flows}
 \end{figure*}
%%%%%%%%%%%%%%%%%%%%%%%%%%%%%%%%%%%%%%%%%%%%%%%%%%%%%%%%%%%%%%%
\section{Preliminaries}
\label{sec:prelim}
%\textbf{Data integration system}
%Semantic integration of big data entails data variety by enabling the resolution of several interoperability conflicts, e.g., structuredness, schematic, completeness, domain, granularity, and entity matching conflicts. These conflicts arise because data sources may have different data models, follow various data representation schemes, and contain complementary information. Furthermore, a real-world entity may be represented using multiple properties or at multiple levels of detail. Thus, data integration techniques able to solve such interoperability issues while addressing data complexity challenges imposed by big data characteristics are demanded. To be able to integrate these sources in a unified way, semantic interoperability conflicts need to be identified~\cite{bellazzi2014big}.

Data Ecosystems 
%(DEs) are data-driven infrastructures that enable stakeholders to exchange and integrate data~\cite{capiello2020data,oliveira2018data}. 
comprise various computational methods to overcome interoperability issues while preserving data privacy, security, and sovereignty. They can be aligned to international data strategies, e.g., the 
European Data Strategy~\cite{EDS}, representing, thus, 
crucial technological building blocks for digitalization and data markets, as well as for enhancing competitiveness and digital sovereignty.
A DE can be centralized, and maintain shared data sources and host services on top of these sources. Moreover, whenever data privacy policies regulate data exchange, a materialized integration of the data is impossible. In this case, several DEs can be interconnected into a DE network ~\cite{capiello2020data}. As an individual DE, each node maintains and exchanges data; it can also perform data management and analytical tasks.
DEs resort to semantic data models for providing a uniform view of heterogeneous data sources.  
Moreover, mapping rules relating to how data sources are defined in terms of the semantic data models are included. Lastly, a DE can also be enhanced with a meta-layer that describes business models, data access regulations, and data exchange contracts.

%\textbf{International Data Spaces}
The International Data Spaces (IDS)~\cite{IDS} is 
an industrial initiative that follows the DE concept. The IDS reference architecture IDS aims at
\begin{inparaenum}[\bf i\upshape)]\item data governance according to regulations imposed by data providers; \item  ensuring a trusted and secure data exchange; 
\item semantically representing main data concepts and relationships; \item exchanging formats and protocols; and \item providing software design principles for guiding the implementation of the reference architecture components.
\end{inparaenum}
IDS provides building blocks for the development of data-driven services, while data sovereignty for data providers is guaranteed.  IDS propose a message-based infrastructure to enable the communication of the different nodes and components in a DE. Moreover, IDS resorts to the Semantic Web standards to express the content and meaning of the shared data source. The Resource Description Framework (RDF) and ontologies defined using RDF is proposed to specify meta-data, and data control and protection in a decentralized or federated DE. The IDS shared information model states standards for representing Content, Concept, Community of Trust, Commodity, and Communication. Proposed W3C standards include SHACL\footnote{\url{https://www.w3.org/TR/shacl/}} are proposed to express content and integrity constraints; SKOS\footnote{\url{https://www.w3.org/2004/02/skos/}} for modeling concepts and relationships; and PROV\footnote{\url{https://www.w3.org/TR/prov-overview/}} for representing data and service provenance.

%%%%%%%%%%%%%%%%%%%%%%%%%%%%%%%%%%%%%%%%%%%%%%%%%%%%%%%%%%%%%%%
\section{Motivating Scenario}
\label{sec:example}
One of the long-term objectives of the EU is creation of common market that will eliminate trade barriers between EU Member States. 
%In the 1990s, the concept of liberalization of the energy sector was introduced to achieve benefits and lower prices for consumers while guaranteeing the security of supply and promoting energy efficiency and renewable energy resources (RES). 
Studies~\cite{ponce2020} have shown that the policy was partially effective at the EU level (being more especially in high-income countries) taking into account the dynamization of the economy and the achieved environmental sustainability. 
The penetration of variable renewable energy sources in the electricity sector increased significantly over the last decade and that allowed the renewable energy suppliers to boost their production and consumption. However, there is still a lack of progress in some countries and overall the EU energy market remains rather fragmented into sub-markets with limited cross-border trade and competition.

\subsection{Electricity Balancing and Commercial Flows on Country Level} 
%Electricity is a commodity that, by its nature, it is difficult to store, however can be sold and traded. At any given moment, the total electricity withdrawals (including  losses) should equal total injections in a control area that is composed of one or several countries. Therefore, on a control block level, a set of actors are responsible for power stability, quality electricity balancing.  
Figure \ref{fig:flows} gives a simplified illustration of electricity flows (blue arrows) and commercial flows (red arrows) between market participants, while in reality, the electricity infrastructure and data exchange processes are very complex, i.e., infrastructure consists of many energy systems (generation, transmission, demand infrastructures). The data sources are related to wind power systems, solar power systems, conventional power plants, cooling, heating, and lighting systems as well as smart grids. They represent measurements in different domains, e.g., energy consumption, energy generation, system outages, failures, weather, and energy transmission. These data sources are characterized by the dominant Big Data dimensions, i.e., volume, velocity, variety, veracity, and value.
Modernization of the grid implies fast integration of RESs, adapted power system planning, new forecasting methods, more flexible use of power plants, standardized data exchange, increased transfer capacity, and others. Additionally, the volatile production of renewable energy sources creates particular challenges for the daily electricity balancing process (i.e., balancing the deviations between the planned or forecast production and demand, on the one side, and the actual performance in real-time, on the other side~\cite{Veen2016}). While the RES installations can be built relatively quickly, the integration occur when the independent producers ensure compliance with grid code requirements~\cite{Brundlinger2019}, as well as, when the basic grid support services are in place. 

%Therefore, the main objective of power system operator (e.g., TSO - the transmission system operator) is to keep the energy supplied by energy service companies (ESCO) in balance with electricity consumption. While on short time scale the goal is to maintain power quality, voltage and grid stability, on medium time scale the scheduled production should meet the planned demand. In order to meet the demand in a reliable and efficient manner, the suppliers have to take into account the variability and the degree of uncertainty of RES power output (independent producers of electricity) and to ensure adequate reserves and sufficient capacities from conventional energy sources. 
Therefore, the integration of distributed variable generation from (independent) producers in the grid is an important subject and therefore it should be adequately addressed. 

\subsection{Data Exchange Requirements}
\begin{figure}[t!]
    \centering
    \includegraphics[width=0.9\columnwidth]{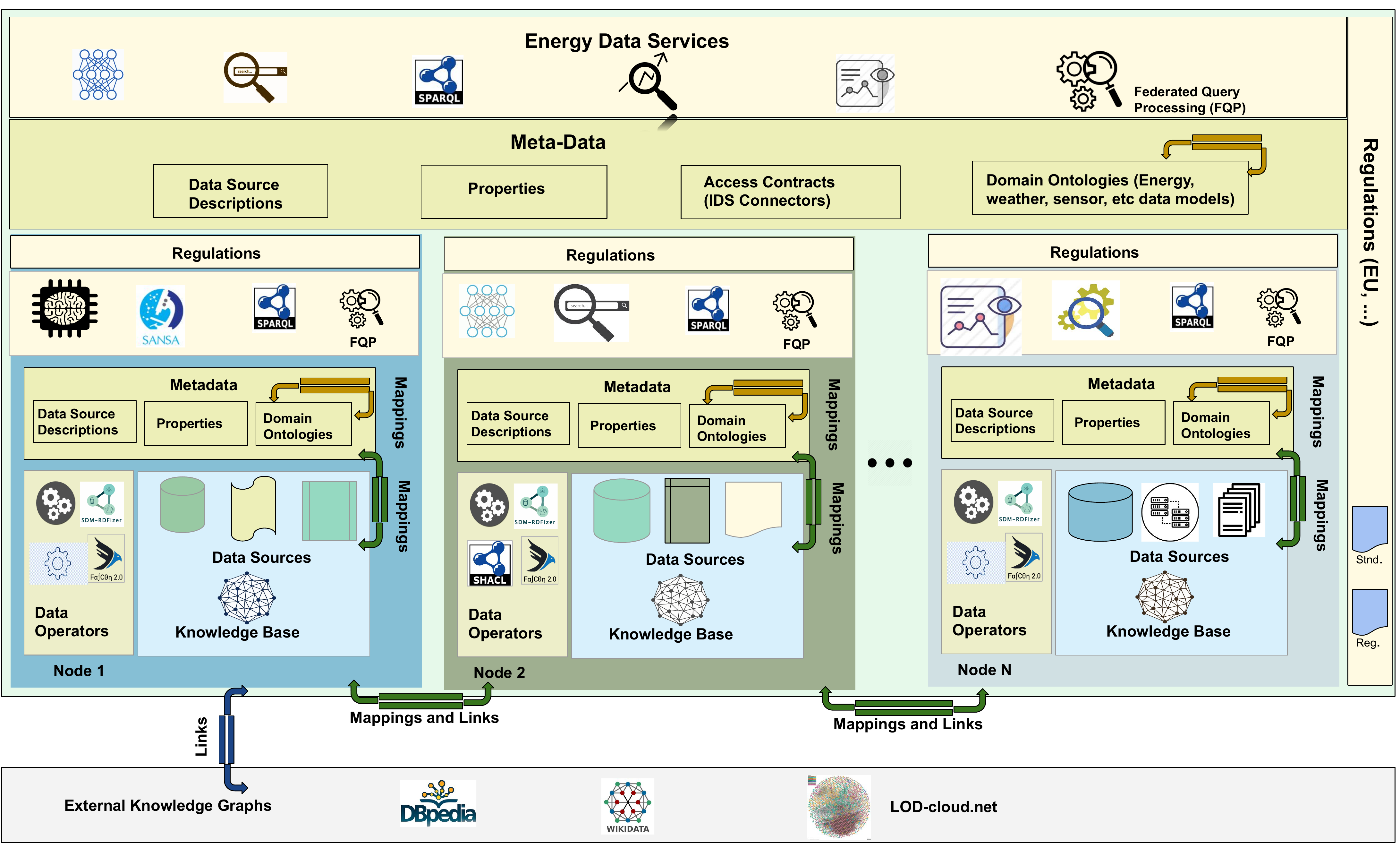}
    \caption{The Energy Big Data integration platform as a Data Ecosystem}
    \label{fig:energydataeES}
\end{figure}

The ability of two or more networks, systems, applications, components, or devices from the same or different vendors to exchange and subsequently use that information to perform required functions is called \textit{interoperability}. \textit{Syntactic interoperability} is the capability of two or more systems of communicating and exchanging data, e.g., specified data formats (e.g., XML), communication protocols (TCP/IP), and the like are fundamental tools of syntactic interoperability.  \textit{Semantic interoperability}~\cite{Nilsson2018} is the ability of systems to exchange information with unambiguous meaning.
 
Regardless of the type of infrastructure (e.g., wind plant, photovoltaic power plant), it is necessary to enable common understanding and messages exchange between different energy stakeholders including raw data (base reading, measurements), processed information (e.g., forecasts, alerts), and market information. The data-driven frameworks previously motivated demand the satisfaction of the following requirements whenever data is exchanged. 

Transmission System Operator (TSO):
\begin{itemize}
    \item \textbf{RQ-1}. For the cross-border electric energy balancing, TSO needs to exchange balancing plans.
    \item \textbf{RQ-2}. In order to plan capacities and electricity cross-border exchanges, TSO receive bids from Balancing Service Provider (BSP) about corresponding volume of balancing energy for the duration of a contract.
    \item \textbf{RQ-3}. TSO collects load information at different points for the grid it operates.  
    \item \textbf{RQ-4}. Based on the balancing contracts, collected bidding information, and collected load, TSO assess the balancing needs and sends plans to BRP for producing  electricity.
\end{itemize}
Balance Service Providers (BSP) and Balance Responsible Party (BRP):
\begin{itemize}
    \item \textbf{RQ-5}. BSP receives bids from producers (BRP) about the expected realization (short, medium and long term) in order to produce a more accurate day-ahead forecast.
    \item  \textbf{RQ-6}. BSP collects different meteorological data in order to plan the energy mix (activation and deactivation of conventional producers).
    \item \textbf{RQ-7}. BRP and BSP prepares (short, medium and long term) forecasts, publish the information in transparent way and sends to TSO. 
    \item \textbf{RQ-8}. BSP and BSP collect infrastructure health information and sends monitoring information to TSO. 
\end{itemize}
%Distribution System Operator (DSO):
%\begin{itemize}
%    \item \textbf{RQ-9}. In order to estimate the effects of integrated Renewable Energy Sources (RES) Producers in the power system, DSO receives information from producers about the quality of produced electricity.
%    \item \textbf{RQ-10}. DSO collects infrastructure health information and sends monitoring information to TSO.
%\end{itemize}

\subsection{Interoperable Analytical Services Requirements}
This subsection presents example scenarios where data-driven methods are required i.e. analytical services to forecast energy consumption and predict maintenance. 

\noindent\textbf{Transmission System Operator (TSO)}: 

\textbf{Load/Demand forecasting}: Electricity demand forecasting is a central and integral process for planning periodical operations and facility expansion in the electricity sector and involves accurate prediction of both magnitudes and geographical locations of electric load over the different periods of the planning horizon. 
%Load forecasting can be divided into three categories: short-term forecasts (from one hour to one week), medium-term forecasts (from a few weeks to a few months and even up to a few years) and long-term forecasts which is a crucial part in the electric power system planning, tariff regulation and energy trading. A long-term forecast is generally known as an annual peak load.
There are several factors that will be taken into consideration for load forecasting, which can be classified as time factor, economic factor, weather condition and customer factor.

\noindent \textbf{Balance Responsible Party}:

\textbf{RES forecasting}: RES 
%producers need a service for more accurate prediction of renewable energy generation. The 
(wind power) forecasting yields estimate the variable power injected in the distribution grid. This allows prediction when the transformer connecting the distribution grid to the transmission grid will be overloaded, i.e., when local wind turbine generator production will be very high. 
%The various forecasting approaches can be classified according to the type of input (weather prediction, wind turbine generators data, historical production data). Statistically based approaches allow very short-term predictions (2 hours). 
One of the key challenges for day-ahead forecasting of wind energy remains unscheduled outages that can have large effects on the forecasts for small systems, while the effect is small on the overall grid. 

\textbf{Predictive maintenance}: The continuous monitoring of asset performance generates input that can be used for predictive analytics and to provide early warnings of component/object failures (e.g., RES plant/component). Identifying problems before they occur helps to reduce unscheduled downtime, improve plant maintenance and optimize asset performance.
%Therefore a service that identify rare events that could occur in power plant infrastructure due to infrastructure health problems, progressive degradation or failure can be deployed. 
%By monitoring the output from the RES power plant using the PMU unit and doing advanced power quality (PQ) analytics close to the source, events can be detected and labeled. By gradually creating a database of events by learning from historical data, one could use this classification to find abnormal functioning of the system before it leads to failure.

\section{Data Ecosystems for Energy Big Data Management and Analytics}
\label{sec:approach}
Herein, we propose an Energy Big Data integration platform as an instantiation of a Data Ecosystem (DE)~\cite{capiello2020data}, see
Figure~\ref{fig:energydataeES}.
%\begin{table*}[t!]
%\centering
%	\caption{Exemplary Scenarios}
%	\label{tab:Scenarios}
 %\begin{tabular}{|p{12.25cm}|} 
 %\hline
%\rowcolor{LightCyan}
%\textbf{TransmissionSystem   Operator (TSO)}: \\
%\hline
% \textbf{Load/Demand forecasting}.  The aims of this use case are load forecasting and prediction of the load pattern. It involves accurate prediction of both magnitudes and geographical locations of electric load over the different periods of the planning horizon. Load forecasting is divided into three categories: short-term forecasts, medium-term forecasts, and long-term forecasts which is a crucial part in the electric power system planning, tariff regulation and energy trading.
 %\\
%\hline
%\hline
%\rowcolor{LightCyan}
%\textbf{Renewable Energy Sources (RES) Producer}: \\
%\hline
%\textbf{RES forecasting}. RES allows prediction of when the transformer connecting the distribution grid to the transmission grid will be overloaded, i.e., when local wind turbine generator production will be very high. The various forecasting approaches can be classified according to the type of input (weather prediction, wind turbine generators data, historical production data). Statistically based approaches allow very short-term predictions (2 hours). \\ 
%\hline
%\hline
%\rowcolor{LightCyan}
%\textbf{Producer}: \\ 
%\hline
%\textbf{Predictive maintenance}. The aim of this use case is to provide services that able to identify rare events that could occur in power plant infrastructure due to infrastructure health problems, progressive degradation or failure.\\ 
% \hline
% \end{tabular}
% \end{table*}
\subsection{Energy Big Data Integration Platform}
\begin{figure*}[ht!]
    \centering
    \includegraphics[width=0.8\columnwidth]{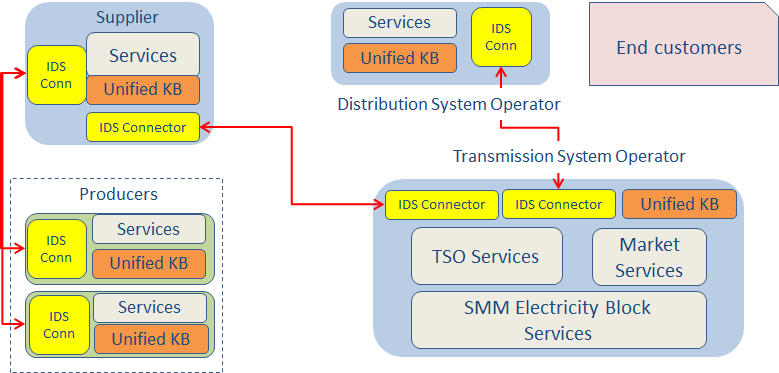}
    \caption{Multi-party data exchange based on IDS concept}
    \label{fig:instantiation}
\end{figure*}
An Energy Big Data integration platform is composed of several data integration platforms (one per Node i). Each node corresponds to a DE and can be integrated on the central level through mappings among nodes, data sharing, and service agreements. Each node (in \autoref{fig:energydataeES} denoted by Node) applies a data integration process on a specific use case and can deploy its services for query processing, analytics as well as dashboards. Communication between nodes needs to be through an access agreement and can employ data connectors (IDS connectors) to secure data exchange according to data access contracts and regulations. Nodes have control over their data and may have data integrated in unified knowledge graphs.  Moreover, each individual knowledge graph can be linked to knowledge graphs in other nodes, or to external knowledge graphs like DBpedia~\cite{AuerBKLCI07}, Wikidata~\cite{VrandecicK14}, or others in the Linked Open Data cloud\footnote{\url{https://lod-cloud.net/}}. Metadata is expressed using common semantic data models (CIM, DCAT, SKOS), and diverse mapping rule languages (e.g., RML or SPARQL) that are utilized in order to define (present) each pilot data sets in terms of the semantic data models. This platform %enables pilots to preserve data sovereignty, privacy, and protection of data and analytical outcomes, as foreseen in IDS. More importantly, it
represents a decentralized infrastructure empowered with the components that pave the way for interoperability across stakeholders. 
\subsection{Instantiating a DE}
The main features of the energy data integration platform are illustrated in the instantiation of a DE; for instance, in the Serbian pilot depicted in Figure~\ref{fig:instantiation}, DEs shall be instantiated at 
\begin{itemize}
    \item Producers site (e.g., at a wind power plant, a unified knowledge graph shall be integrated with the production forecast and the predictive maintenance services);
    \item Supplier site, an organization that integrate data from many producers and sell electricity to TSO (e.g., the Power Industry of Serbia might be interested to integrate the data sources from power plants it owns and manages);
    \item Transmission System Operator site, an organization that operates and balances the grid (e.g., the Joint Stock Company EMS might be interested in improving the data integration and the transparency of data exchanged with other actors).  
\end{itemize}

For instance at the DE of Transmission System Operator four main data sources are currently available for integration as follows i) the Joint Stock Company (JSC) EMS Transparency platform\footnote{\url{https://transparency.ems.rs/}}; ii) ENTSO-E Transparency platform\footnote{\url{https://transparency.entsoe.eu}}; iii) Meteorological data from WeatherBit\footnote{\url{https://www.weatherbit.io}}; and iv) data from SCADA system (archive data for RES production and aggregated load)\footnote{\url{http://www.pupin.rs/en/products-services/process-management/scada/}}. 

%SCADA RES data is available in real time through a MySQL database. 
Data operators for preprocessing, mapping, linking, transformation, and validation are applied to the pilot data sources for creating a materialized version of the unified knowledge graph. The mappings between data sources and the target ontology are part of the DE as well. Furthermore, mappings between concepts from different ontologies are part of each DE. Data sources are also described in terms of provenance and main properties; these descriptions are utilized for the creation of a knowledge graph (e.g., by using  SDM-RDFizer \cite{iglesias2020sdm}) and during query processing (e.g., by using Ontario \cite{EndrisRVA19}). Links between entities in knowledge graph and external data sources can be done by performing entity linking. Tools like Falcon2.0~\cite{SakorSPV20} can be applied to linking the pilots' datasets with external knowledge graphs like DBpedia and Wikidata, while SHACL validation engines (e.g., Trav-SHACL~\cite{FRV2021}) enable the validation of integrity constraints. Lastly, RDF knowledge graph will feed the Semantic based analytics engine SANSA~\cite{LehmannSBWSEBCS17} to perform tasks of knowledge discovery and prediction.

%%%%%%%%%%%%%%%%%%%%%%%%%%%%%%%%%%%%%%%%%%%%%%%%%%%%%%%%%%%%%%%
\section{Application: A Use Case}
\label{sec:proof} 
%Analysis of Renewable Energy Potential
Although Serbia is not an EU country, the Energy Sector Development Strategy is based on the EU Energy Roadmap; one of the goals is to increase the RES share. Also, the information about the quantity of energy produced from RES has to be presented to the end user (guarantee of origin) with a document issued by the Distribution System Operator. Hence, the information from the producers, via suppliers and TSO, shall be available to end user (from Serbia and/or abroad) (Figure~\ref{fig:instantiation}).
%Once the IDS certificates have been exchanged between the sending and receiving parties, trustful connections are established that allow standardized data access and information retrieval not just on country level, but on a level of participants in the IDS compliant digital ecosystem. 

\subsection{Developing a Global Schema for the Energy Domain }
%Different ontologies are proposed in the literature for development of a global schema including (i) Upper ontologies (e.g., SUMO, Dolce, BFO), (ii) Core ontologies (e.g., Agent Ontology, Time Ontology), (iii) Domain ontologies for a specific domain and (iv) Domain-specific ontologies that can be reused and extended in order to meet a specific need of the application. 
%In the literature review phase, we concentrated on gathering information about the common semantic concepts and properties applicable for the targeted scenarios.
%, see also Table~\ref{tab:Scenarios}
For development of a global schema different existing data models have been consulted and considered for reuse such as
\begin{itemize}
    \item the IEC Common Information Model standards (CIM)\footnote{\url{https://www.dmtf.org/standards/cim/cim\_schema\_v2530)}};
    \item the Smart Appliances REFerence ontology (SAREF);
    \item the IDS\footnote{\url{https://international-data-spaces-association.github.io/InformationModel/docs/index.html}} Information Model; 
\item the SEAS - Smart Energy Aware Systems\footnote{\url{https://w3id.org/seas/}}.
\end{itemize}
\begin{figure*}[t!]
    \centering
    \includegraphics[width=0.9\columnwidth]{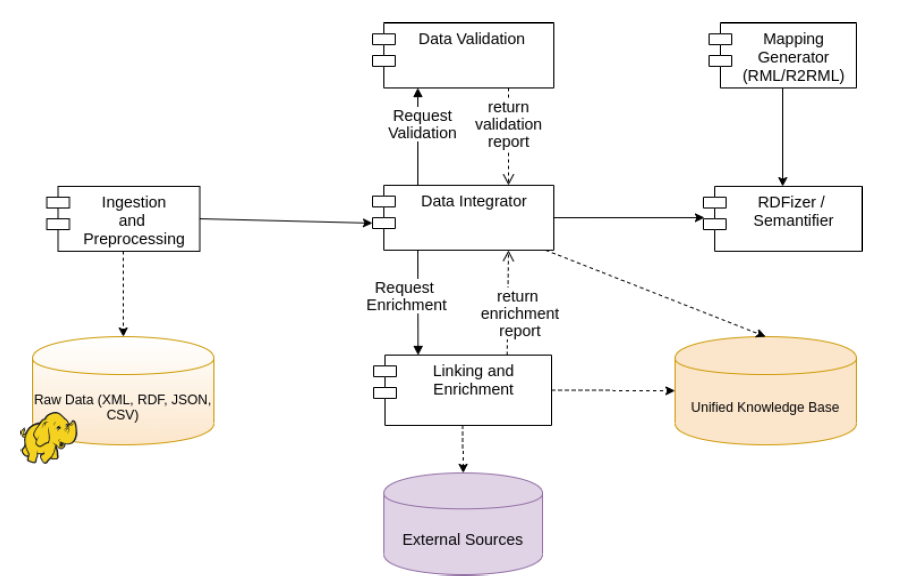}
    \caption{Unified Knowledge Graph Creation Process}
    \label{fig:unifiedKB}
\end{figure*}

The selection has been done based on a set of scenarios (electricity balancing services, predictive maintenance services, services for residential, commercial and industrial sector). In our analysis, we have used the semantic CIM model\footnote{\url{  https://ontology.tno.nl/IEC_CIM/}}. 
It is a canonical taxonomy in the form of packages of UML class diagrams referring to the components of power utility networks with functional definitions and measurement types to a high degree of granularity (packages: Core, Topology, Wires, Generation, LoadModel, Outage, SCADA, ControlArea and others). 
The concepts selected for reused come from different packages. For instance, \texttt{cim:PowerSystemResource} (Core package) can be an item of equipment such as a Switch, a \texttt{cim:EquipmentContainer} containing many individual items of equipment such as a Substation. Each \texttt{cim:PowerSystemResource} is registered on the grid (\texttt{cim:RegisteredResource}) and belong to a control area (\texttt{cim:HostControlArea}) that is operated by a Contro Area Operator. The \texttt{cim:ControlAreaOperator} is responsible for stabilizing the system frequency (\texttt{cim:Frequency}); it is therefore also called frequency control. 
\\\textbf{Example - Load/Demand forecasting}: 
The system is balanced by utilizing both supply and demand resources. However, the existing electric power systems were not initially designed to incorporate different kinds of generation technology (\texttt{cim:Plant}) in the scale that is required today. Historically, balancing the system has been maintained mostly by directing thermal power plants to increase or reduce output (\texttt{cim:ActivePower}) in line with changes in demand. 
%With significant penetration of distributed generation, the distribution network has become an active system with power flows and voltages determined by the generation and by the loads (see \texttt{cim:AreaLoadCurve}). To forecast the load for the next period \texttt{cim:ControlAreaOperator} needs a service for load prediction for different \texttt{cim:LoadForecastType}. The service is semantically described with the SEAS Forecast Ontology. Considering that \texttt{cim:ActivePower} and \texttt{cim:AreaLoadCurve}, the Transmission System Operator (TSO) (\texttt{cim:ControlAreaOperator}) participates on the \texttt{cim:EnergyMarket}; it serves as 
%\texttt{cim:TradeResponsibleParty} and \texttt{cim:ImbalanceSettlementResponsible}. Further, together with other competing entities, it submits offers (\texttt{cim:Bid}) of power and energy to meet the next day's load.
%Hence, \texttt{cim:ControlAreaOperator} has to prepare \texttt{cim:BidSelfSced} for the \texttt{cim:HostControlArea}. The scheduling activity is performed on regular bases, e.g. each hour (\texttt{cim:Schedule}). 

\textbf{Example - RES forecasting}:
%Another use case is related to a resource connected to the Grid. Independent producers (IPP) and producers (\texttt{cim:Producer}) from distributed and renewable sources (DER) will be actors in the balance reserve market in the future. The goal of this scenario is to develop and test a service for more accurate prediction of renewable energy generation from RES plants (\texttt{cim:Plant}). 
Electricity production however from solar and wind plants (\texttt{cim:Plant}) is subject to considerable forecast errors that drive demand for balancing, i.e., for ( \texttt{cim:ReserveReq}). The amount for each reservation is defined by the Agreement (\texttt{cim:Agreement}) on the provision of system services signed between the transmission system operator and the balancing service provider (\texttt{cim:BalanceSupplier}).

Once the global schema has been developed, it can be used across the nodes established in the energy data ecosystem.

\subsection{Unified Knowledge Graph Creation Process}
In this section, 
%two scenarios of the knowledge graph creation process 
%and their pros and cons are discussed.
%Creating a knowledge graph from heterogeneous data sources, for example at the Supplier site, requires the description of the entities in the data sources using RDF vocabularies, as well as the performance of curation and integration tasks to reduce data quality issues, e.g., missing values or duplicates.
two types of knowledge graph creation strategies are discussed: materialized (i.e., data warehousing) and virtual (i.e., Data Lake). Both strategies are applicable for the use cases discussed above. 
\\
\textbf{Materialized Knowledge Graph Creation Process:}
In a materialized knowledge graph creation process, data from individual data sources are loaded and materialized into an RDF format and stored in a physical database, the so-called triplestore. 
Figure~\ref{fig:unifiedKB} shows the data curation and integration sub-components for creating a unified knowledge graph. %The ingestion and preprocessing component is the gateway to the knowledge graph creation process.
Input data from Producers' data sources is first stored in a raw data repository, i.e., staging repository. Any preprocessing steps, such as cleaning, normalization, and aggregation, that are predefined for input data are applied and provenance is recorded. The data integrator component then orchestrates the knowledge graph creation process according to the data source’s configuration by invoking the Linking and Enrichment, RDFizer/Semantifier, and Data Validation sub-components and finally integrating data to the Supplier's unified knowledge graph. The Linking and Enrichment component performs entity linking and enrichment using external as well as existing materialized knowledge graphs. The RDFizer/Semantifier component transforms non-semantic, i.e., raw, data to RDF graph based on mapping rules. Data validation component checks data constraint conformance. 
\\
\textbf{Virtual Knowledge Graph Creation Process:}
In a virtual knowledge graph creation process, data remains in the sources (in raw format) and is accessed as needed during query time. 
%The federated query processing component can handle this process. 
The federated query processing component employs the data source descriptions stored in the metadata store to perform the integration during query time. Metadata about the number of data sources available, the provenance of the data sets, and mapping rules to transform data to RDF graph are stored in a separate data store available for both materialized and virtual data integration processes. 
%If the data sets are already included in the materialized knowledge graph, then the federated query processing component can directly access them without performing data transformation at query time. However, if the data sources are stored in raw format, then the data transformation rules will be applied only for the part of the data set required to answer the query. 
%Figure~\ref{fig:federatedQP} shows the basic components of the virtual knowledge graph creation process through a federation system. 
The federated query engine will use SPARQL query language\footnote{\url{https://www.w3.org/TR/rdf-sparql-query/}} to access the unified knowledge graph, as described in the next Section.
%\begin{figure*}[t!]
%    \centering
%    \includegraphics[width=1.4\columnwidth]{Fig05.png}
%    \caption{Federated Query Processing at TSO level}
%    \label{fig:federatedQP}
%\end{figure*}

\subsection{Traversing the Knowledge Graph}
 Once the knowledge graph creation process is established, exploring the knowledge base will be possible via a query engine. 
 %Additionally, data exploration and knowledge discovery services can be employed. Results of executing a federated query can be used as input of Data Analytics or Knowledge Discovery tasks.
%As the knowledge base is defined through mapping to semantic data models for energy, the query processing engine is able to process queries posed using the SPARQL query language.
If the materialization approach is applied and data is stored in a centralized triple store, e.g., Virtuoso, then the knowledge base can be accessed using SPARQL query over the query engine embedded in the triple store. However, if the size (in terms of volume) of the materialized knowledge base is big, then partitioning and distribution is necessary for timely response from the query engine and handling the resource requirements to store such large data in expensive servers. 
%Such distribution of data needs to be accessed through a federated query engine that is able to distribute the posed query to each partition and merge data returned from them. 
%Virtual integration approach can also be applied over heterogeneous data sources. In this case, the query processing engine not only query each data source and merge results but also should be able to transform raw data to the semantic models specified in the mappings during query time. 

\subsection{Federated Query Processing}
Federated query processing system provides a unified access interface to a set of autonomous, distributed, and heterogeneous data sources. While distributed query processing systems have control over each data set, federated query processing engines have no control over data sets in the federation, and data providers can join or leave the federation at any time and modify their data sets independently. 
%Query Processing in the context of data sources in a federation is more difficult than centralized systems because of the different parameters involved that affect the query processing engine's performance. Data sources in a federation might contain data fragments about an entity, have different processing capabilities, and support different access patterns, access methods, and operators. 
The role of federated query processing engines is to transform a query, i.e., the federated query, expressed in terms of the global schema into an equivalent query expressed in the schema of the data sources, i.e., the local query. The local query represents the federated query's actual execution plan by the federation's data sources. An essential part of query processing in the context of federated data sources is query optimization. 
%Since many execution plans are correct transformations of the same federated query, the one that optimizes (minimize) resource consumption should be retained. The performance of query processors can be measured by the total cost that will be used in processing the query and the response time of the query, i.e., the time elapsed for executing the query.

\textbf{Example:} Let us consider the following question expressed in SPARQL: “A list of countries, their renewable energy plants, and respective installed generation capacity for the year 2020”
%\begin{lstlisting}[label=lst:sparql1]
\begin{verbatim}
PREFIX wd:     <http://www.wikidata.org/entity/>
PREFIX wdt:    <http://www.wikidata.org/prop/direct/>
PREFIX energy: <http://w3id.org/energy/>

SELECT DISTINCT ?country ?productionType ?measure
WHERE {
?genCapacity    a  energy:GenerationCapacity .
?genCapacity    energy:productionType ?productionType .
?genCapacity    energy:country        ?country .
?genCapacity    energy:measure        ?g_measure .
?genCapacity    energy:agg_year       "2020" .
?productionType wdt:P279              wd:Q12705 .
} 

\end{verbatim}
%\end{lstlisting}
To execute this query, a federation of knowledge graphs is needed, i.e., the Energy DE knowledge graph and the external knowledge graphs like Wikidata \footnote{\url{https://www.wikidata.org/}}. The federated query engine maintains metadata about these knowledge graphs, and it is able to select them as relevant sources for the query. Then, once the knowledge graphs are selected, the federated query engine decomposes the query into subqueries SQ1 and SQ2, and executes them over the selected knowledge graphs, respectively. Query SQ1 is defined as follows; it is executed against the local knowledge graph.  

%\begin{lstlisting}[label=lst:sparql2]
\begin{verbatim}
PREFIX energy: <http://w3id.org/energy/>
SELECT DISTINCT ?country ?productionType ?measure
WHERE {
?genCapacity    a  energy:GenerationCapacity .
?genCapacity    energy:productionType ?productionType .
?genCapacity    energy:country        ?country .
?genCapacity    energy:measure        ?g_measure .
?genCapacity    energy:agg_year       "2020" .
} 
\end{verbatim}
%\end{lstlisting}

On the other hand, query SQ2 is defined and evaluated over Wikidata.
%\begin{lstlisting}[label=lst:sparql3]
\begin{verbatim}
PREFIX wd:  <http://www.wikidata.org/entity/>
PREFIX wdt: <http://www.wikidata.org/prop/direct/>

SELECT DISTINCT ?productionType
WHERE {
	?productionType     wdt:P279  wd:Q12705 .
} 
\end{verbatim}
%\end{lstlisting}

%Executing query SQ1 produces instantiantions for the variables $?country$ $?productionType$, and $?measure$. While executing SQ2 over the Wikidata knowledge graph produces answers describing renewable energy plants; these plants are described as values of the variables $?productionType$ and $?productionTypeLabel$.

The federated query engine performs join against the results produced by the execution of queries SQ1 and SQ12; the values of the variable $?productionType$ are utilized as a join column. As a result, only one renewable energy plant can be matched, i.e., the wind power. It is important to highlight that without the integration of the transparency platform data and the linking of the corresponding production types with Wikidata, this query could not be executed. 

 \section{Discussion}
\label{sec:discussion}
%In the last decade, the Big Data paradigm has gain momentum and is generally employed by businesses on a large scale to create value that surpasses the investment and maintenance costs of data. 
The energy sector is an example where tremendous amounts of data are collected from numerous sensors, which are generally attached to different plant subsystems. The new paradigm of DEs for smart grids that includes renewable energy sources challenges the existing network infrastructure and the energy management systems even more. DEs address the following challenges:
\begin{itemize}
    \item definition of new approaches to data management and processing and extending the service portfolio of various energy stakeholders in order to achieve two-way flows of electricity;
    \item deploying distributed/edge processing and data analytics technologies to optimize the operation of the real-time energy system management and automate the “monitor-forecast-optimize-control” loop;
    \item implementation of effective integration of relevant digital technologies for transforming the system from top-down centralized production and rigid distribution framework to collaborative ecosystem of self-managed prosumers able to act independently on the liberalized energy markets.
\end{itemize}

%To achieve the targets envisioned in the latest EU energy strategy and the European Green Deal Action Plan~\cite{GreenDeal}, a standard framework is needed to encapsulate, communicate and manage the distributed assets in the energy value chain. 
In this paper, we showcased how a “networks” of distributed data integration platforms can be instantiated in the energy value chain for establishing a "network of trusted data". Some benefits for main actors are
\begin{itemize}
\item \textbf{Secure data exchange}: Using the IDS concept that features various levels of protection, data is exchanged securely across the entire data supply chain (and not just in bilateral data exchange). 
\item \textbf{Data governance and sovereignty}: Data owner determines the terms and conditions of use of the data provided, while data sovereignty always remains with the respective Data Provider. The Provider makes data available to be requested by certain contractors in the Data Space by its own rules. 
%Additionally, Provider can offer data services  (e.g. via an »AppStore«) to be found by all DE participants.
\item \textbf{Innovative scalable and replicable energy management services}: The Data Spaces opens opportunities for new data-driven and model-driven services that will complement and enhance the existing e.g. balancing services, energy generation and consumption intelligent forecasts services, energy performance assessment services, etc.  
\end{itemize}
 
 \section{Related Work}
\label{sec:related}
Gelhaar and Otto~\cite{GelhaarO20} highlight the value of data-driven solutions in the digitization era and outline the challenges that need to be addressed in DEs in emerging areas like maritime, manufacturing, and science. Controlled and secured data exchange in a traceable way are among the most relevant challenges. 
% As shown in the described scenarios, DEs for energy big data are demanded to provide computational methods and semantic-based formalisms (e.g., ontologies) to represent the meaning of the data to be shared and processed. The meta-data layer comprises  unified schemas, mappings between data sets and concepts in the unified schema, and alignments across ontologies. Furthermore, following the IDS reference architecture, 
% integrity constraints are represented using declarative formalisms (e.g., SHACL), while data provenance and quality is described based on standard vocabularies (e.g., PROV and DQV). These semantic descriptions provide building blocks for documenting data sharing, integration, and processing. As a result, services for tracking down DE components can be provided.   
 
Several approaches have been defined to follow the DE architecture with the aim of solving interoperability across heterogeneous data sets during query processing time; they are usually named as federated query engines. Exemplary approaches include GEMMS~\cite{QuixHV16}, PolyWeb~\cite{Khan2019}, BigDAWG~\cite{Duggan:2015:BPS:2814710.2814713}, 
Constance~\cite{HaiGQ16}, and Ontario~\cite{EndrisRVA19}. 
These systems collect metadata about the main characteristics of their data sets, e.g., formats and query capabilities. 
Additionally, they resort to a global ontology to describe contextual information and 
the relationships among data sets, for purposes of optimized data integration, query processing, and automated schema discovery in quasi-central settings. 
Metadata have shown to be crucial for enabling these systems to perform query processing effectively. Knowledge-driven DEs are built on these results and make available the semantic description of the data collections made available by stakeholders. 
Furthermore, a DE empowers federated query processing engines with factual statements about the integrity constraints satisfied by the data retrieved and merged during query processing. As a consequence, a new paradigm shift in data management is devised towards tracing down data integration during query processing.

\section{Conclusion and Future Work}
\label{sec:conclusion}
%are cyber-physical energy systems, the next evolution step of the traditional power grid and are characterized by a bidirectional flow of information and energy. 
One of the requirements related to data access procedures in Smart Grids and future electricity markets is related to interoperability of energy services. Therefore, the overall goal of the paper is to showcase and evaluate Data Ecosystems and the IDS concept for the energy sector. 
%The IDS initiative is based on the use of semantic technologies for creation of knowledge-based systems that will aid machines in integrating and processing resources contextually and intelligently. 
In our work, we showed how DEs provide the building blocks for enhancing the interoperability of energy management applications/services; they also enable the integration of energy data in the European Energy Data Space. The meta-data layer in DEs together with the internal SCADA information model can be used as an information hub (‘knowledge graphs’) for (1) building data connectors that will facilitate integration of services in future integrated energy systems and (2) improving the explainability of machine learning services / analytical applications. The selection of models has been done based on a set of scenarios (electricity balancing services, predictive maintenance services, services for residential, commercial and industrial sector). 
%The proposed approach is being used in the EU-funded H2020 project PLATOON. The development of all the computational components and unified schemas to fulfil the requirements of each pilot is part of our future agenda. 
\section*{Acknowledgements}
This work has been partially supported by the EU H2020 funded projects PLATOON (GA No. 872592), the EU project LAMBDA (GA No. 809965), and partly by the Ministry of Science and Technological Development of the Republic of Serbia (No. 451-03-9/2021-14/200034) and the Science Fund of the Republic of Serbia (Artemis, No.6527051).

\printbibliography
\end{document}